\documentclass[submission,copyright,creativecommons]{eptcs}
\providecommand{\event}{FROM 2022} 
\usepackage[T1]{fontenc}
\usepackage{graphicx}
\usepackage{todonotes}
\usepackage{caption}
\usepackage{hyperref}
\usepackage{parcolumns}
\usepackage[numbers,sort&compress]{natbib}
\usepackage{ebproof}

\newcommand{\defined}[1]{\lceil #1 \rceil} 
\newcommand{\total}[1]{\lfloor #1 \rfloor} 

\newcommand{\brac}[1]{\llbracket #1 \rrbracket}

\newcommand{\DEFINEDNESS}{\mathsf{DEF}}
\newcommand{\NAIVESORTS}{\mathsf{NAIVESORTS}}
\newcommand{\SORTS}{\mathsf{SORTS}}

\usepackage[utf8]{inputenc}
\usepackage{listings}
\usepackage[frozencache]{minted}
\usepackage{ifplatform}
\input{lstcoq.sty}

\usepackage{newfloat}
\usepackage{amsmath,amssymb,mathtools,amsthm}
\usepackage{stmaryrd}
\usepackage{multirow}
\usepackage{cleveref}
\usepackage{tcolorbox}
\usepackage{fancyvrb}
\usepackage{multicol}
\usepackage{prftree}
\usepackage{bm}
\usepackage{ulem}
\usepackage{xspace}
\usepackage{caption}
\usepackage{subcaption}
\usepackage{hyperref}
\newcommand*{\textlabel}[2]{
  \edef\@currentlabel{#1}
  \phantomsection
  #1\label{#2}
}

\usepackage{color}

\tcbuselibrary{listings}

\newcommand{\SVar}{\mathsf{SVar}}
\newcommand{\EVar}{\mathsf{EVar}}
\newcommand{\ee}{\uline}
\newcommand{\se}[1]{\uuline{#1}}
\newcommand{\ld}{\,.\,}
\newcommand{\ev}[2]{|#1|_{#2}}
\newcommand{\NN}{\mathbb{N}}
\newcommand{\apM}{\mathbin{@_M}}
\newcommand{\opene}{\mathsf{open}_\mathsf{ele}}
\newcommand{\opens}{\mathsf{open}_\mathsf{set}}
\newcommand{\lfp}{\mathbf{lfp}}
\newcommand{\FF}{\mathcal{F}}
\newcommand{\prule}[1]{(#1)}
\newcommand{\FV}[1]{FV(#1)}
\newcommand{\hole}{\square}
\newcommand{\K}{$\mathbb{K}$\xspace}

\DeclareFloatingEnvironment[fileext=frm,placement={!ht},name=Frame]{specification}
\newcommand{\Smbs}{{\textsf{Symbol}}}
\newcommand{\Axms}{{\textsf{Axiom}}}
\newcommand{\Sugar}{{\textsf{Notation}}}

\newcommand{\Impts}{{\textsf{Import}}}
\newcommand{\axname}[1]{\textnormal{(\textsc{#1})}}

\newtheorem{definition}{Definition}
\newtheorem{theorem}{Theorem}

\definecolor{light-gray}{gray}{0.95}

\lstdefinelanguage{AML}
{morekeywords={spec, endspec, model, endmodel, of},
backgroundcolor=\color{light-gray},
sensitive=false,
basicstyle=\small,
morecomment=[l][\textit]{//},
morecomment=[s]{/*}{*/},
keywordstyle=\bfseries\sffamily,
morestring=[s]""
}

\tcbset{size=small}

\ifwindows
  \newenvironment{coqcode}
  {\begingroup\lstset{language=Coq,escapeinside=??}\VerbatimOut{blocklisting-tmp.txt}}
  {\endVerbatimOut\begin{tcolorbox}\lstinputlisting{blocklisting-tmp.txt}\end{tcolorbox}\endgroup}
    
  \newcommand{\coqinline}[1]{\lstinline[language=Coq,mathescape=true,escapeinside=!!]\##1\#}
\else
  \newminted[coqcodebasic]{Coq}{escapeinside=??,mathescape=true,fontsize=\footnotesize}
  \newenvironment{coqcode}
  {\VerbatimEnvironment\begin{tcolorbox}[colback=black!5!white] 
   \VerbatimEnvironment\begin{coqcodebasic}}
  {\end{coqcodebasic}
   \end{tcolorbox}
  }
  
  \newcommand{\coqinline}[1]{\mintinline[escapeinside=??,fontsize=\footnotesize]{Coq}{#1}}
\fi

\begin{document}

\title{Mechanizing Matching Logic In Coq}


\def\titlerunning{Mechanizing Matching Logic In Coq}
%
\author{P\'{e}ter Bereczky\institute{E\"{o}tv\"{o}s Lor\'{a}nd University, Hungary} \and
Xiaohong Chen\institute{University of Illinois Urbana-Champaign, USA} \and
D\'{a}niel Horp\'{a}csi\institute{E\"{o}tv\"{o}s Lor\'{a}nd University, Hungary} \and
Lucas Pe\~{n}a\institute{University of Illinois Urbana-Champaign, USA} \and
Jan Tu\v{s}il\institute{Masaryk University, Brno, Czech Republic}
}

\def\authorrunning{P. Bereczky et al.}
%
%
\maketitle              

%

\begin{abstract}
Matching logic is a formalism for specifying, and reasoning about, mathematical structures, using patterns and pattern matching. Growing in popularity, it has been used to define many logical systems such as separation logic with recursive definitions and linear temporal logic. In addition, it serves as the logical foundation of the K semantic framework, which was used to build practical verifiers for a number of real-world languages. Despite being a fundamental formal system accommodating substantial theories, matching logic lacks a general-purpose, machine-checked formalization. Hence, we formalize matching logic using the Coq proof assistant. Specifically, we create a new representation of matching logic that uses a locally nameless encoding, and we formalize the syntax, semantics, and proof system of this representation in the Coq proof assistant. Crucially, we prove the soundness of the formalized proof system and provide a means to carry out interactive matching logic reasoning in Coq. We believe this work provides a previously unexplored avenue for reasoning about matching logic, its models, and the proof system.

\end{abstract}


\section{Introduction}




Matching logic~\cite{Rosu17,ChenRosu19Mml}
is a unifying logical 
framework for defining the formal semantics of programming languages
and specifying their language tools. 
Given a programming language $L$,
its formal semantics is defined by a \emph{matching logic theory} $\Gamma^L$,
i.e., a set of axioms.
Many language tools, such as parsers, interpreters, compilers, and even deductive
program verifiers, are best-effort implementation 
of matching logic reasoning. 
The correctness of language tools is justified by matching logic proof objects,
checkable by a 240-line proof checker~\cite{ml-checker}. 

The formal semantics of many real-world programming languages 
have been \emph{completely defined} as matching logic theories.
These languages include
C~\cite{HER15}, 
Java~\cite{k-java},
JavaScript~\cite{k-js},
Python~\cite{python-semantics}, 
Rust~\cite{krust-singapore,krust-shanghai},
Ethereum bytecode~\cite{HSZ+18},
x86-64~\cite{DPK+19}, 
and LLVM~\cite{k-llvm}.
The \K framework (\url{https://kframework.org})
is a best-effort implementation of matching logic.
From the formal semantics of these real-world languages,
\K automatically generates implementations and formal analysis tools,
some of which have been commercialized~\cite{rv-match}.
Ultimately, \K can be used as a tool for formally defining languages,
and as a tool for formally reasoning about properties of programming languages and programs.

\K currently provides the most comprehensive support for 
\emph{automated reasoning} for matching logic
by means of various algorithms that are specifically targeted at
the (automatic) generation of language tools
such as interpreters and deductive verifiers, which, respectively,
\emph{are} specific forms of matching logic reasoning,
and an integration with the state-of-the-art SMT solvers such as 
Z3~\cite{DMB08}.
However, there is no ``exit solution'' in \K when
these automatic algorithms and external solvers fail,
in which case \emph{lemmas}, whose correctness
is justified externally, often informally,
need to be manually added to fill in the reasoning gaps.

In this paper, we aim to bridge the aforementioned reasoning gaps
by bringing \emph{interactive reasoning} to matching logic. 
Specifically, we give, for the first time, a complete mechanization 
of matching logic in Coq~\cite{CoqArt}.






\subsection*{Contributions}

In this work, we investigate the formal definition of matching logic in an interactive theorem prover, with the aim of enabling computer-aided reasoning \emph{about} and \emph{within} matching logic:
\begin{itemize}
\item \emph{Mechanize the soundness of the matching logic proof system}---while matching logic has been proven sound on paper, its soundness has not been formalized yet. This soundness proof is crucial in providing the highest level of assurance for mechanized reasoning.
\item \emph{Enable interactive, mechanically verified reasoning about matching logic theories}---the formalization needs to leverage the full power of the theorem prover to encode matching logic theories in a modular way and do reasoning at the highest abstraction level possible.
\end{itemize}

Following upon similar formalization efforts (such as the solutions to the POPLMark Challenge~\cite{poplmark}), we have decided to encode the logic in a widely used and mature system, the Coq proof assistant~\cite{CoqArt}. Furthermore, we carry out the embedding in a locally nameless~\cite{ChargueraudLocallyNameless,Leroy07alocally,McBrideM04,Gordon93} representation, which, in contrast to named approaches, is more amenable to computer-aided verification. The entire formalization is open, and available online~\cite{ml-formalization}. The paper shows the following main contributions:
\begin{itemize}\setlength\itemsep{0em}
\item A locally nameless definition of the matching logic syntax, semantics, and proof system;
\item The design of the embedding of the locally nameless matching logic in the Coq proof assistant;
\item The first \emph{mechanized} proof of the soundness of the matching logic proof system;
\item Example theories with proofs about their semantic and proof-theoretical properties in the mechanized matching logic;
\item A preliminary implementation of a matching logic proof mode to simplify interactive reasoning.
\end{itemize}
%
%
%
%
%
The paper is structured as follows.
Section~\ref{sec:ml-intro} introduces matching logic, in
a locally nameless representation. In Section~\ref{sec:formalization}, we outline the Coq
formalization, including some technical challenges faced.
Section~\ref{sec:soundness} discusses examples of meta-level reasoning and the soundness proof in particular.
Section~\ref{sec:proofmode} presents an example interactive proof in our formalization.
Finally, Section~\ref{sec:related-work} discusses related
work, and
Section~\ref{sec:conclusion} presents areas for future work and
concludes.



\section{Introduction to Matching Logic}\label{sec:ml-intro}

In this section, we present the syntax, semantics, and proof
system of matching logic. 
Unlike in previous work \cite{ChenRosu19Mml,ChenRosu20Binders,ChenLucanuRosuInitAlgebraTR},
here we introduce a
\emph{locally nameless} presentation of matching logic.
This new presentation is more convenient to be formalized in proof assistants,
which is discussed in detail in \Cref{sec:formalization}.

\subsection{Matching Logic Locally Nameless Syntax}
\label{sec:matching-logic-syntax}

The syntax we present in this section is in literature known
as that of a \emph{locally nameless} one~\cite{ChargueraudLocallyNameless,Leroy07alocally,McBrideM04,Gordon93}.
A locally nameless syntax is a combination of the traditional
named representation and the entirely nameless de Bruijn encoding,
using named variables if they occur free and de Bruijn encoding
if they are bound.
In particular, it distinguishes free and bound variables on the syntactic level,
enabling capture-avoiding substitutions without variable renaming.
This design decision eliminates the need for reasoning about $\alpha$-equivalence.

Firstly, we define matching logic \emph{signatures}.
A signature provides us infinitely many named variables and a set of
(user-defined) constant symbols. 

\begin{definition}[Signatures]\label{def:ml-signature}
A \emph{matching logic signature} is a tuple
$(\EVar,\SVar,\Sigma)$ where
\begin{itemize}
\item $\EVar$ is a set of \emph{element variables},
denoted $x$, $y$, \dots;
\item $\SVar$ is a set of \emph{set variables},
denoted $X$, $Y$, \dots;
\item $\Sigma$ is a set of \emph{(constant) symbols}, denoted
$\sigma$, $f$, $g$, \dots
\end{itemize}
Both $\EVar$ and $\SVar$ are countably infinite sets.
$\Sigma$ is a countable set, possibly empty. 
When the sets of variables are understood from the context,
we feel free to use $\Sigma$ to denote a matching logic signature. 
\end{definition}

Given a signature $\Sigma$, the syntax of matching logic
generates a set of well-formed formulas, called \emph{patterns}.
In the following, we first define pseudo-patterns. 

\begin{definition}[Locally Nameless Representation of Pseudo-Patterns]\label{def:ml-pseudopattern}
Given a signature $\Sigma$, the set of \emph{pseudo-patterns}
is inductively defined by the following grammar:
$$
\varphi \Coloneqq 
x \mid X \mid \ee{n} \mid \se{N} \mid \sigma \mid
\varphi_1 \, \varphi_2 \mid \bot \mid \varphi_1 \to \varphi_2
\mid \exists \ld \varphi \mid \mu \ld \varphi
$$
\noindent
In the above grammar, 
$x$ and $X$ are element and set variables, respectively;
$\ee{n}$ and $\se{N}$ are de Bruijn indices that represent
the bound element and set variables, respectively,
where $n,N \in \NN$;
$\sigma$ is any symbol in the given signature $\Sigma$;
$\varphi_1 \, \varphi_2$ is called \emph{application},
where $\varphi_1$ is applied to $\varphi_2$;
$\bot$ and $\varphi_1 \to \varphi_2$ are propositional operations;
$\exists \ld \varphi$ is the FOL (first-order logic)-style quantification;
and $\mu \ld \varphi$ is the least fixpoint pattern.
Both $\exists$ and $\mu$
use the nameless de Bruijn encoding for their bound variables.
\end{definition}

\begin{definition}[Locally Nameless Representation of Patterns]\label{def:ml-pattern}
We say that a pseudo-pattern $\psi$ is \emph{well-formed},
if (1) for any subpattern $\mu \ld \varphi$, the nameless variable bound by $\mu$ has no negative occurrences in $\varphi$; and
(2) all of its nameless variables (i.e., de Bruijn indices)
are correctly bound by the quantifiers $\exists$ and $\mu$, that is,
for any de Bruijn indices $n$ and $N$ that occur in $\psi$,
\begin{itemize}
\item $\ee{n}$ is in the scope of at least $n+1$ $\exists$-quantifiers;
\item $\se{N}$ is in the scope of at least $N+1$ $\mu$-quantifiers.
\end{itemize}
A well-formed pseudo-pattern is called a \emph{pattern}. For example, the definition of transitive closure in \Cref{sec:example_eq} is a pattern, but $\exists \ld \ee{0} \to \ee{1}$ is not, since $\ee{1}$ is not bound by any quantifier.
\end{definition}

\paragraph{Opening quantified patterns.}
In a locally nameless representation, we use both
named and unnamed variables.
Named variables are for free variables,
while unnamed variables are for bound variables. 
Therefore, when we have a pattern $\exists \ld \varphi$
(similarly for $\mu \ld \varphi$) and want to extract its body,
we need to assign a fresh named variable to the unnamed variable
that corresponds to the topmost quantifier $\exists$ (resp. $\mu$).
This operation is called \emph{opening} a (quantified) pattern~\cite{ChargueraudLocallyNameless}.
We use $\opene(\varphi,x)$ and $\opens(\varphi,X)$
to denote the opening of the bodies of the quantified patterns
$\exists \ld \varphi$ and $\mu \ld \varphi$, respectively,
where $x$ and $X$ are the corresponding new named variables.

These opening operations are special instances of the general case of substitution,
which is defined in the usual way\footnote{We denote substitution with $\phi[\psi/x]$, where $x$ (a \textit{bound} or \textit{free}, \textit{set} or \textit{element} variable) is replaced by $\psi$ in $\phi$.}.

\subsection{Matching Logic Models and Semantics}
\label{sec:matching-logic-semantics}

In this section, we formally define the models and semantics of matching logic. 
Intuitively speaking, matching logic has a \emph{pattern matching semantics}.
A matching logic pattern is interpreted as the \emph{set}
of elements that \emph{match} it. 

Firstly, we define the notion of matching logic models.

\begin{definition}[Matching Logic Models]\label{def:ml-model}
Let $\Sigma$ be a signature.
A \emph{$\Sigma$-model}, or simply a \emph{model},
is a tuple $(M, @_M, \{\sigma_M\}_{\sigma \in \Sigma})$
where:
\begin{itemize}
\item $M$ is a nonempty carrier set;
\item $@_M \colon M \times M \to \mathcal{P}(M)$ is a binary application function,
where $\mathcal{P}(M)$ denotes the powerset of $M$;
\item $\sigma_M \subseteq M$ is the interpretation of $\sigma$,
for each $\sigma \in \Sigma$.
\end{itemize}
We use the same letter $M$ to denote the model defined as above.
\end{definition}

Next, we define valuations of variables.
Note that matching logic has both element and set variables.
As expected, a valuation assigns element variables to elements
and set variables to subsets of the underlying carrier set. 
Formally,

\begin{definition}[Variable Valuations]\label{def:ml-valuation}
Given a signature $\Sigma$ and a $\Sigma$-model $M$,
a variable valuation $\rho$ is a mapping such that:
\begin{itemize}
\item $\rho(x) \in M$ for all $x \in \EVar$;
\item $\rho(X) \subseteq M$ for all $X \in \SVar$.
\end{itemize}
\end{definition}

We are now ready to define the semantics of matching
logic patterns. 

\begin{definition}[Matching Logic Semantics]\label{def:ml-semantics}
Given a matching logic model $M$ and a variable valuation $\rho$,
we define the semantics of any (well-formed) pattern $\psi$,
written $\ev{\psi}{M,\rho}$, inductively as follows:
\vspace*{-0.4cm}
\begin{table}[h]
\renewcommand{\arraystretch}{1.5}
\begin{tabular}{c c}
$\ev{x}{M,\rho} = \{ \rho(x) \}$; &
$\ev{X}{M,\rho} = \rho(X)$; \\
$\ev{\sigma}{M,\rho} = \sigma_M$; &
$\ev{\bot}{M,\rho} = \emptyset$; \\
$\ev{\varphi_1 \to \varphi_2}{M,\rho} = M \setminus ( \ev{\varphi_1}{M,\rho} \setminus \ev{\varphi_2}{M,\rho})$; &
\hspace{1cm}
$\ev{\varphi_1 \, \varphi_2}{M,\rho} = \bigcup_{a_1 \in \ev{\varphi_1}{M,\rho}}\bigcup_{a_2 \in \ev{\varphi_2}{M,\rho}}
a_1 \apM a_2 $; \\
\multicolumn{2}{c}{$\ev{\exists \ld \varphi}{M,\rho}
   = \bigcup_{a \in M} \ev{\opene(\varphi,x)}{M,\rho[a/x]}$ for fresh $x \in \EVar$;} \\
\multicolumn{2}{c}{$\ev{\mu \ld \varphi}{M,\rho} = \lfp \ \FF^\rho_{\varphi,X}$,
where $\FF^\rho_{\varphi,X}(A) = \ev{\opens(\varphi,X)}{M,\rho[A/X]}$
for fresh $X \in \SVar$.}
\end{tabular}
\end{table}

\noindent The above definition is well-defined ($\opene$ and $\opens$ are defined at the end of
\Cref{sec:matching-logic-syntax}). For examples, we refer to the formalization~\cite{ml-formalization} and to~\cite[Section~4]{ml-explained}.
\end{definition}

In the following, we define validity and the semantic entailment relation
in matching logic. 

\begin{definition}\label{def:ml-satisfaction}
For $M$ and $\varphi$, we write $M \models \varphi$ iff
$\ev{\varphi}{M,\rho} = M$ for all valuations $\rho$. 
For a pattern set $\Gamma$, called a \emph{theory}, we write $M \models \Gamma$ iff
$M \models \varphi$ for all $\varphi \in \Gamma$. 
We write $\Gamma \models \varphi$ iff for any $M$,
$M \models \Gamma$ implies $M \models \varphi$. 
\end{definition}

\subsection{Matching Logic Proof System}

In this section, we present the proof system of matching logic.
Matching logic has a Hilbert-style proof system with
19 simple proof rules, making it small and easy to implement. 
The proof system defines the \emph{provability relation}
written as $\Gamma \vdash \varphi$, where
$\Gamma$ is a set of patterns
(often called a \emph{theory} and the patterns are called \emph{axioms})
and $\varphi$ is a pattern that is said to be \emph{provable}
from the axioms in $\Gamma$.

\begin{table*}[t]
\renewcommand{\arraystretch}{1.35}
\caption{Matching Logic Proof System under Locally Nameless Representation\\
($C_1,C_2$ are application contexts, $FV(\varphi)$ denotes the set of free variables in $\varphi$)}\label{tab:ps}
\begin{tabular}{llll} 
\hline
\emph{\textbf{Proof Rule Names}} & \emph{\textbf{Proof Rules}} &
\emph{\textbf{Proof Rule Names}} & \emph{\textbf{Proof Rules}} 
\\ 
\hline  
\prule{Proposition 1} & $\varphi_1 \to (\varphi_2 \to \varphi_1)$ &
\prule{Proposition 2} & 
 \renewcommand{\arraystretch}{1}
 \begin{tabular}{@{}l@{}}
 $(\varphi_1 \to (\varphi_2 \to \varphi_3)) \to$\\\kern1em$(\varphi_1 \to \varphi_2)\to (\varphi_1 \to \varphi_3)$ 
 \end{tabular}\\[15pt]
\prule{Proposition 3} & $((\varphi \to \bot) \to \bot) \to \varphi$ &
\prule{Modus Ponens} & 
{
\begin{prooftree}
\hypo{\varphi_1}
\hypo{\varphi_1 \to \varphi_2}
\infer2{\varphi_2}
\end{prooftree}
} \\
\prule{$\exists$-Quantifier} &
  $\opene(\varphi,x) \to \exists \ld \varphi$ with $x \in \EVar$ \kern-8em\\[5pt]
\prule{$\exists$-Generalization} &
{
\begin{prooftree}
\hypo{\opene(\varphi_1,x) \to \varphi_2}
\infer1[with $x \not\in \FV{\varphi_2}$]{(\exists \ld \varphi_1) \to \varphi_2}
\end{prooftree}
}  \kern-8em \\[10pt]
\hline
\prule{Propagation Left$_\bot$} & $\bot \, \varphi \to \bot$ 
& \prule{Propagation Right$_\bot$} & $\varphi \, \bot \to \bot$  \\
\prule{Propagation Left$_\vee$} & $(\varphi_1 \vee \varphi_2) \, \varphi_3 \to (\varphi_1 \, \varphi_3) \vee (\varphi_2 \, \varphi_3)$ \kern-10em \\
\prule{Propagation Right$_\vee$} & $\varphi_1 \, (\varphi_2 \lor \varphi_3) \to (\varphi_1 \, \varphi_2) \vee (\varphi_1 \, \varphi_3)$ \kern-10em \\
\prule{Propagation Left$_\exists$} 
& $ (\exists \ld \varphi_1) \, \varphi_2 \to \exists \ld \varphi_1 \, \varphi_2$ &
\prule{Propagation Right$_\exists$} 
& $ \varphi_1 \, (\exists \ld \varphi_2) \to \exists \ld \varphi_1 \, \varphi_2$ \\[3pt]
\prule{Framing Left} & 
{
\begin{prooftree}
\hypo{\varphi_1 \to \varphi_2}
\infer1{\varphi_1 \, \varphi_3 \to \varphi_2 \, \varphi_3}
\end{prooftree}
}
&
\prule{Framing Right} & 
{
\begin{prooftree}
\hypo{\varphi_2 \to \varphi_3}
\infer1{\varphi_1 \, \varphi_2 \to \varphi_1 \, \varphi_3}
\end{prooftree}
}
\\[8pt]
\hline
\prule{Substitution} &
{
\begin{prooftree}
\hypo{\varphi}
\infer1{\varphi[\psi/X]}
\end{prooftree}
}
& \prule{Pre-Fixpoint} &\kern-1em
$\varphi[(\mu \ld \varphi) / \se{0}]\to \mu \ld \varphi$ \\[10pt]
\prule{Knaster-Tarski} &
{
\begin{prooftree}[rule margin=0pt]
\hypo{\varphi_1[\varphi_2/\se{0}] \to \varphi_2}
\infer1{(\mu \ld \varphi_1) \to \varphi_2}
\end{prooftree}
}
\\[8pt]
\hline
\prule{Existence} & $\exists \ld \ee{0}$ &
\prule{Singleton} & 
\kern-2em$\neg(C_1[x \land \varphi] \wedge C_2[x \land \neg \varphi])$
\\\hline
\end{tabular}
\renewcommand{\arraystretch}{1}
\end{table*}

We present the proof system of matching logic 
in Table~\ref{tab:ps}.
To understand it, we first need to define the notion of
\emph{contexts} and a particular type of contexts called
\emph{application contexts}. 

\begin{definition}
A \emph{context} $C$ is simply a pattern with one unique placeholder denoted $\hole$. 
We write $C[\varphi]$ to mean the result of plugging $\varphi$ in $\hole$ in the context $C$. 
We call $C$ an \emph{application context} if from the root of $C$ to $\hole$
there are only applications; that is, $C$ is (inductively) constructed as follows:
\begin{itemize}
\item $C$ is $\hole$ itself, called the identity context; or
\item $C\equiv C_1 \, \varphi$, where $C_1$ is an application context; or
\item $C\equiv\varphi \, C_2$, where $C_2$ is an application context.
\end{itemize}
\end{definition}

\noindent
The proof rules in Table~\ref{tab:ps} can be divided into four categories:
\begin{itemize} 
\item \textbf{FOL reasoning} containing the standard proof rules as in FOL;
\item \textbf{Frame reasoning} consisting of six propagation rules and two framing rules, which allow us to propagate formal reasoning that is carried out within an application context throughout the context. Note that we proved these rules equivalent to the ones in~\cite{ChenRosu19Mml}, where application contexts are not splitted into applications to the left and right;
\item \textbf{Fixpoint reasoning} containing the fixpoint rules as in modal $\mu$-calculus;
\item \textbf{Miscellaneous rules}.
\end{itemize}

Next, we state the soundness theorem of the proof system, which has been proved by induction over the structure of the proof $\Gamma \vdash \varphi$ in~\cite{ChenRosu19Mml}. We elaborate on the mechanization of this proof in \Cref{sec:soundness}.

\begin{theorem}[Soundness Theorem]
$\Gamma \vdash \varphi$ implies $\Gamma \models \varphi$. 
\end{theorem} 


\subsection{Example Matching Logic Theories}

\paragraph{First-order logic.}

It is easy to see that matching logic is at least as expressive as classical first-order logic, and Chen et al.~\cite{ml-explained} show that it is indeed more expressive than FOL. At the same time, they describe a direct and natural connection between FOL and matching logic.

A FOL term $t$ is interpreted as an element in the underlying carrier set.
From the matching logic point of view, $t$ is a pattern
that is matched by \emph{exactly one element}. 
We use the terminology \emph{functional patterns} to refer to
patterns whose valuations are singleton sets.
Intuitively, FOL terms \emph{are} functional patterns in matching logic. 
%
FOL formulas are interpreted as two logical
values: true and false.
In matching logic, there is a 	simple analogy
where we use the empty set $\emptyset$ to represent the logical false
and the total carrier set to represent the logical true.
A pattern whose interpretation is always $\emptyset$ or the full set
is called a \emph{predicate pattern}.
Intuitively, FOL formulas \emph{are} predicate patterns in matching logic.


\paragraph{Equality.}
\label{sec:example_eq}

%
Although matching logic has no built-in notion of equality, it can be easily defined using a construct called \emph{definedness}.
Formally, we define $\Sigma^\DEFINEDNESS=\{ \defined{\_} \}$, $\Gamma^\DEFINEDNESS=\{ \defined{x} \}$ ($\defined{x}$ denotes $\defined{\_}\, x$);
this axiom ensures that whenever a pattern $\varphi$ is matched by some model element, the pattern $\defined{\varphi}$ is matched by all elements of the model, and vice versa.
Equality is then defined as a notation $\varphi_1 = \varphi_2 \equiv \neg \defined{\neg (\varphi_1 \leftrightarrow \varphi_2 )}$,
intuitively saying that there is no element that would match only one of the two formulas.
It is easy to see that for any pattern $\varphi$, the pattern $\defined{\varphi}$ is a predicate pattern, and that equality of two patterns is a predicate.
With this, we can similarly define other notations, such as membership, subset, totality, etc. seen below.
To make things easier, we use the notion of matching logic \emph{specification} introduced in~\cite{ml-explained}.
The signature $\Sigma^\DEFINEDNESS$ and theory $\Gamma^\DEFINEDNESS$ are then defined by Spec.~\ref{spec:definedness}. For more details on definedness, we refer to \Cref{Sec:formequality}.
\begin{specification}[th]
\begin{tcolorbox}
\begin{lstlisting}[mathescape=true,language=AML]
spec $\DEFINEDNESS$
  $\Smbs$ $\defined{\_}$
  $\Sugar$
    $\arraycolsep=1.0pt\begin{array}{rclrcl}
       \defined{\varphi} &\equiv& \defined{\_} \, \varphi&
       \quad \total{\varphi} &\equiv& \neg \defined{ \neg \varphi} \\
       \varphi_1 = \varphi_2 &\equiv& \total{\varphi_1 \leftrightarrow 
       \varphi_2} &
       \quad \varphi_1 \neq \varphi_2 &\equiv& \neg (\varphi_1 = \varphi_2) \\
       x \in \varphi &\equiv& \defined{x \wedge \varphi}&
       \quad \varphi_1 \subseteq \varphi_2 &\equiv& \total{\varphi_1 \rightarrow
       \varphi_2} \\
       x \not\in \varphi &\equiv& \neg(x \in \varphi)&
       \quad \varphi_1 \not\subseteq \varphi_2 &\equiv& \neg(\varphi_1 
          \subseteq \varphi_2)
    \end{array}$
  $\Axms$ 
    $\axname{Definedness}$ $\defined{x}$
endspec
\end{lstlisting}
\end{tcolorbox}
\caption{Definedness and related notions}
\label{spec:definedness}
\end{specification}

\vspace{-2ex}
\paragraph{Induction and transitive closure.}

Chen et al. show how matching logic, by using the application and least fixpoint operators, can not only axiomatize equality, but also product types ($\langle \_ , \_ \rangle$), and inductive types~\cite{ml-explained}.
Hence, another notable example of matching logic's expressiveness is that it can specify the transitive closure of a binary relation $R$ the following way (where $\in$ is defined by Spec.~\ref{spec:definedness}):
%
$$
\mu X \ld R \lor \exists x \ld \exists y \ld \exists z \ld \langle x,z \rangle \land \langle x, y \rangle \in X \land \langle y, z \rangle \in X
$$
\noindent
Or rather, the same matching logic pattern expressed in the locally nameless representation:
$$
\mu \ld R \lor \exists \ld \exists \ld \exists \ld \langle \ee{2},\ee{0} \rangle \land \langle \ee{2}, \ee{1} \rangle \in \se{0} \land \langle \ee{1}, \ee{0} \rangle \in \se{0}
$$

\section{Matching Logic Formalization in Coq}\label{sec:formalization}
In this section, we describe how the locally nameless matching logic (as defined in \Cref{sec:ml-intro}) has been encoded in Coq.
The formalization is distributed as a library including the definition of the
logic as well as that of some standard theories, with the two dependencies being the Std++ library~\cite{stdpp} and the Equations plugin~\cite{coq-equations}.
Some of our proofs rely on classical and extensionality axioms (namely, functional extensionality, propositional extensionality,
and the axiom of excluded middle), but these are known to be compatible with Coq's logic~\cite{carneiro_master}.

We implement matching logic in a \emph{deep embedding} style; that is, formulas, models, and proofs are represented as data in Coq.
Presumably, a shallow embedding could provide a more lightweight implementation, but deep embedding has couple of advantages for our use cases; mainly, it allows us to inspect matching logic proofs without reflection, reason about them directly (for instance, when checking the side conditions of the \emph{deduction theorem} from~\cite{chen-rosu-2019-tr-mml}) and to extract them from Coq (see \Cref{sec:conclusion})
\footnote{It is beyond the scope of this work to tell if a shallow embedding would be more suitable for object-level reasoning.}
.

\subsection{Syntax}
We represent a matching logic signature $(\EVar,\SVar,\Sigma)$,
defined in \Cref{def:ml-signature}, as an instance of the \coqinline{Class Signature} that encapsulates the sets of variables and the set of symbols. The sets for variables are required to be countable and infinite, and in addition, it is also required that equality on variables and symbols is decidable.
We also provide an off-the-shelf, string-based instance of the type class as a default option.
Although this instance will suffice for most applications, to prove the completeness of the matching logic proof system, a more general carrier set will be needed\footnote{We refer to the proof of the completeness of matching logic without $\mu$ and the extension lemma~\cite{chen-rosu-2019-tr-mml}.}.
The way we represent free variable names has some features in common with the concept of \emph{atoms} used in \emph{nominal} approaches~\cite{nominalcoq} (e.g., any countably infinite set with decidable equality can be used for names), but in the locally nameless approach we do not rely on permutative renaming when implementing capture-avoidance.

We formalize (pseudo-)patterns (Definition~\ref{def:ml-pseudopattern}) as the inductive definition
\coqinline{Inductive Pattern : Set}.
Due to using a locally nameless embedding, 
we have separate constructors for free variables (named
variables) and bound variables (de Bruijn indices).
The main advantage this provides is in the equivalence
of quantified patterns. Specifically, in a fully named representation,
the patterns $\exists x \ld x$ and $\exists y \ld y$ are equivalent,
yet not syntactically equal.
To formally prove their equivalence, we would need a notion of
$\alpha$-equivalence of patterns that is constantly applied.
However, in our locally nameless representation, both these can only
be represented by the pattern $\exists \ld \ee{0}$, and thus are
syntactically equal, so no additional notions of equality need to be
supplied.
This is implicitly used throughout our soundness proof and in proving
properties of specifications.

%

\vspace*{-2ex}

\paragraph{Well-formedness.}\label{sec:wf}
For practical reasons, the type \coqinline{Pattern} corresponds to the definition
of pseudo-patterns, while the restrictions on non-negativity and scoping defined for patterns (\Cref{def:ml-pattern}) are
implemented as a pair of \coqinline{bool}-valued auxiliary functions:
\begin{coqcode}
well_formed_positive : Pattern -> bool
well_formed_closed   : Pattern -> bool
\end{coqcode}
\noindent where the first function performs a check for the positivity constraint and the second one checks the scoping requirements. The predicate \coqinline{well_formed_closed} is constructed from two parts:
\begin{coqcode}
well_formed_closed_ex_aux : Pattern -> nat -> bool
well_formed_closed_mu_aux : Pattern -> nat -> bool
\end{coqcode}
\noindent
These predicates return \coqinline{true}, when the parameter pattern contains only smaller unbound de Bruijn indices for element and set variables than the given number.
A pattern is \coqinline{well_formed} if it satisfies both \coqinline{well_formed_closed} and \coqinline{well_formed_positive}.
Most of our functions operate on the type \lstinline{Pattern} without the well-formedness constraint;
we use the constraint mainly for theorems.
This way we separate proofs from data.



\paragraph{Substitution and opening.}
In the locally nameless representation, there are separate substitution functions for bound and free variables (both for element and set variables). In our formalization, we followed the footsteps of Leroy~\cite{Leroy07alocally}, so the bound variable substitution decrements the indices of those (bound) variables that are greater than the substituted index.
We define
\begin{coqcode}
Definition evar_open (k : db_index) (x : evar) (p : Pattern) : Pattern.
Definition svar_open (k : db_index) (X : svar) (p : Pattern) : Pattern.
\end{coqcode}
\noindent
which correspond to $\opene(\varphi,x)$ and
$\opens(\varphi,X)$ from Section~\ref{sec:matching-logic-syntax}, with a difference that this version of opening allows for
substitution of any de Bruijn index, not only the one corresponding
to the topmost quantifier (\ee{0} or \se{0}).

\paragraph{Derived notations.}

Matching logic is intentionally minimal.
As a consequence, any non-trivial theory is likely to heavily rely on notations that abbreviate common operations.
Besides basic notations for boolean operations, universal quantification and greatest fixpoint,
one also can define equality, subset and membership relations on top of the definedness symbol (as we did in Spec.~\ref{spec:definedness}),
which is not part of matching logic, but is usually considered as a part of
the ``standard library'' for the core logic.

Coq provides (at least) two ways to extend the syntax of the core logic with derived notations.
The first option is to use Coq's \coqinline{Notation} mechanism. For example, the following would define the notations for negation, disjunction, and conjunction:
\begin{coqcode}
Notation "! p" := (p ---> ?$\bot$?).
Notation "p or q" := (! p ---> q).
Notation "p and q" := ! (! p or ! q).
\end{coqcode}
The problem with this is that the pattern
\begin{coqcode}
(x ---> ?$\bot$?) ---> ?$\bot$?
\end{coqcode}
\noindent could be interpreted either as \coqinline{! (! x)}, or as \coqinline{x or ?$\bot$?},
which would be confusing to the user, having no control on which interpretation Coq chooses to display.

Therefore, we decided to opt for the other option, representing each derived notation as a Coq \coqinline{Definition}, as in the following snippet:
\begin{coqcode}
Definition patt_not p := p ---> ?$\bot$?.
Definition patt_or p q := patt_not p ---> q.
Definition patt_and p q := patt_not (patt_or (patt_not p) (patt_not q))
\end{coqcode}
We can define the notations on top of these definitions. This way, the user can fold/unfold derived notations as needed.
However, this representation of notations poses another problem: many functions, especially the substitutions 
such as \coqinline{bevar_subst}, preserve the structure of the given formula, but since they build a \coqinline{Pattern}
from the low-level primitives, the information about derived notations is lost whenever such function is called.
We solve this by defining for each kind of syntactical construct (e.g., for unary operation, binary operation, element variable binder)
a type class containing a rewriting lemma for \coqinline{bevar_subst} such as this one:
\begin{coqcode}
Class Binary (binary : Pattern -> Pattern -> Pattern) :=
{
  binary_bevar_subst :
    forall ?$\psi$?, well_formed_closed ?$\psi$? ->
      forall n ?$\phi_1$? ?$\phi_2$?,
        bevar_subst (binary ?$\phi_1$? ?$\phi_2$?) ?$\psi$? n
        = binary (bevar_subst ?$\phi_1$? ?$\psi$? n) (bevar_subst ?$\phi_2$? ?$\psi$? n) ;
(* ... *)
}.
\end{coqcode}
\noindent The user of our library then can instantiate the class for their derived operations
and use the tactic \coqinline{simpl_bevar_subst} to simplify the expressions containing \coqinline{bevar_subst} and \coqinline{evar_open}.



\paragraph{Fresh variables.}
We say that a variable is fresh in a pattern $\varphi$ if it does not occur among the free variables of $\varphi$.
Sometimes (for example, in the semantics of existential and fixpoint patterns) it is necessary
to find a variable that does not occur in the given pattern.
We required the variable types (\coqinline{evar} and \coqinline{svar}) to be infinite, thus we can use the solution of the Coq Std++ library~\cite{stdpp} for fresh variable generation.
We then have a function \coqinline{fresh_evar : Pattern -> evar} and the following lemma:
\begin{coqcode}
Lemma set_evar_fresh_is_fresh ?$\varphi$? : fresh_evar ?$\varphi$? ?$\notin$? free_evars ?$\varphi$?.
\end{coqcode}

\subsection{Semantics}\label{sec:semantics}

On the semantics side
we have a \coqinline{Record Model}.
We do not require the carrier set of the model to have decidable equality. 
We represent the variable valuation function defined in \Cref{def:ml-valuation} as a record of two separate functions, one mapping element variables to domain elements
and another mapping set variables to sets of domain elements.
With this, we define the interpretation of patterns ($\ev{\_}{M,\rho}$ in
\Cref{def:ml-semantics}) as expected, with two points worth noting:
\begin{enumerate}
\item The interpretation of patterns cannot be defined using structural recursion on the formula,
      because in the $\mu$ (and $\exists$) case,
      one calls \coqinline{eval} on \coqinline{svar_open 0 X p'} (and \coqinline{evar_open 0 x p'}), which is not a structural subformula
      of \coqinline{p}. Therefore, we do recursion over the size of the formula,
      implemented as an \coqinline{Equation}:
\begin{coqcode}
Equations eval (?$\rho$? : @Valuation M) (?$\varphi$? : Pattern) : propset (@Domain M) by wf (size ?$\varphi$?) :=
    (* ... *)
    eval ?$\rho$? (patt_mu ?$\varphi'$?) := let X := fresh_svar ?$\varphi'$? in
      @LeastFixpointOf _ OS L (fun S =>
        let ?$\rho'$? := (update_svar_val X S ?$\rho$?) in
          eval ?$\rho'$? (svar_open 0 X ?$\varphi'$?)).
\end{coqcode}

\item We decided to give semantics to patterns that are not well-formed, including arbitrary $\mu$ patterns.
      This way, we do not have to supply the \coqinline{eval} function
      with the well-formedness constraint, which makes it easier to use.
      We did that by
      (1) defining the function \coqinline{LeastFixpointOf} to return
      the intersection of all prefixpoints;
      (2) mechanizing the relevant part of the Knaster-Tarski fixpoint theorem~\cite{tarski1955}, and
      (3) proving that the function associated to a well-formed $\mu$ pattern is monotone.
\end{enumerate}
%

%

\subsection{Proof System}

We formalize the proof system of matching logic as an inductive definition:
\begin{coqcode}
Inductive ML_proof_system theory : Pattern -> Set := (* ... *) .
\end{coqcode}
The proof system is defined as expected; however, one may ask why the proof system lives in \coqinline{Set}
and not in \coqinline{Prop}.
The answer is that in our \emph{deep embedding} we care about the internal structure of matching logic proofs, not only about provability.
We do not want two matching logic proofs to be considered identical only because they prove the same formula,
for at least two reasons.
First, some theorems (e.g. the deduction theorem from~\cite{chen-rosu-2019-tr-mml}) can only be applied to a proof if that proof satisfies particular
conditions regarding its internal structure.
Second, when extracting OCaml or Haskell programs from this Coq formalization,
we need the manipulation of the proof system terms to be preserved;
that will allow us to extract Metamath proof objects in the future (see \Cref{sec:conclusion}).

Another point worth mentioning is that the rules in the formalization of the proof system
often require some well-formedness constraints.
A consequence of this is that only well-formed patterns are provable:
\begin{coqcode}
Lemma proved_impl_wf ?$\Gamma$? ?$\phi$?: ?$\Gamma$? ?$\vdash$? ?$\phi$? -> well_formed ?$\phi$?.
\end{coqcode}

We discuss the soundness of the proof system in
Section~\ref{sec:soundness}.
Crucially, we rely on the $\mu$ patterns to be interpreted as
least fixpoints, as explained in Section~\ref{sec:semantics}.



\section{Reasoning about Matching Logic}\label{sec:soundness}

After encoding matching logic in Coq, we overview some results it allows for concerning meta-level reasoning. In particular, we highlight some challenges we faced when proving the mechanized matching logic proof system sound, and we demonstrate semantics-based reasoning about the theory of equality.

\subsection{Soundness of The Proof System}

The most crucial result of the mechanization of matching logic is the proof of the soundness of its proof system (\Cref{tab:ps}). Even though this theorem has already been investigated in related publications, we have developed the first complete, machine-checked proof, which verifies the prior paper-based results.

We state our soundness theorem below:
\begin{coqcode}
Theorem Soundness : forall phi : Pattern, forall theory : Theory,
  well_formed phi -> theory ?$\vdash$? phi -> theory ?$\vDash$? phi.
\end{coqcode}

The proof of soundness begins via induction on the hypothesis
\coqinline{theory ?$\vdash$? phi}, meaning we consider all cases from
the proof system which may have produced this hypothesis.
Many cases, such as the propositional proof rules, were straightforwardly discharged
using set reasoning. Other cases, like Modus Ponens, were discharged via
the application of the induction hypothesis.

%
The most involved cases were the proof rules involving quantification,
specifically the $\exists$-Quantifier, Prefixpoint, and
Knaster-Tarski rules.
For these rules, the key steps were to establish complex substitution lemmas (separate
lemmas for existential quantification and for $\mu$-quantification).
The proofs of these lemmas were very involved, and we note that the set substitution lemma was not proved in related work previously. For existential quantification, we adapt the
``element substitution lemma'' which appears
in~\cite[Lemma 41]{chen-rosu-2019-tr-mml}.

For the soundness of the Pre-fixpoint and Knaster-Tarski rules with $\mu$-quantification,
we introduce a new similar lemma called
\emph{set substitution lemma}, which
links syntactic substitution with semantic substitution,
stating that the
following two ways of plugging a pattern $\varphi_2$ into a pattern
$\varphi_1$ are equivalent:
\begin{enumerate}
  \item syntactically substitute $\varphi_2$ for a free set variable $X$ in $\varphi_1$ and interpret the resulting pattern;
  \item interpret $\varphi_2$ separately, then interpret $\varphi_1$ in a valuation where $X$ is mapped to the value of $\varphi_2$.
\end{enumerate}

\subsection{Theory of Equality}\label{Sec:formequality}

The formalization also allows for reasoning about matching logic models. We implemented the theory of definedness and equality as presented in~\cite{ml-explained,Rosu17}, and \Cref{sec:example_eq}. Then, we established some results about models that satisfy the definedness axiom, which provides support for common cases of semantic reasoning. This branch of the development showcases applications of our mechanization for reasoning about matching logic models.

\paragraph{Definedness and totality.}

Definedness has the important property that, applied to any formula $\varphi$ which matches at least one model element, the result matches all elements of the model (represented by $\top$):
\begin{coqcode}
Lemma definedness_not_empty_iff : forall (M : @Model ?$\Sigma$?),
    M ?$\vDash^T$? theory ->
    forall (?$\phi$? : Pattern) (?$\rho_e$? : @EVarVal ?$\Sigma$? M) (?$\rho_s$? : @SVarVal ?$\Sigma$? M),
      (@eval ?$\Sigma$? M ?$\rho_e$? ?$\rho_s$? ?$\phi$?) <> ?$\emptyset$? <-> (@eval ?$\Sigma$? M ?$\rho_e$? ?$\rho_s$? ?$\lceil$? ?$\phi$? ?$\rceil$? ) = ?$\top$?.
\end{coqcode}
This is why it is called \emph{definedness}: $\lceil \phi \rceil $ evaluates to full set if and only if $\phi$ is \emph{defined};
that is, matched by at least one element.
Note that one needs the definedness axiom only for the ``if'' part;
the ``only if'' part is guaranteed by the definition of the extension of application:
anything applied to the empty set results in the empty set.
The dual of definedness is called \emph{totality}: a pattern $\phi$ is considered \emph{total} iff it is matched by all elements of the model, and totality of a pattern ($\lfloor \phi \rfloor$) \emph{holds} (is matched by all elements of the model) only in that case:
\begin{coqcode}
Lemma totality_not_full_iff : forall (M : @Model ?$\Sigma$?),
    M ?$\vDash^T$? theory ->
    forall (?$\phi$? : Pattern) (?$\rho_e$? : @EVarVal ?$\Sigma$? M) (?$\rho_s$? : @SVarVal ?$\Sigma$? M),
      @eval ?$\Sigma$? M ?$\rho_e$? ?$\rho_s$? ?$\phi$? <> ?$\top$? <-> @eval ?$\Sigma$? M ?$\rho_e$? ?$\rho_s$? ?$\lfloor$? ?$\phi$? ?$\rfloor$? = ?$\emptyset$?.
\end{coqcode}

\paragraph{Equality.}

As we have seen in \Cref{sec:ml-intro}, equality is defined using totality.
We proved that equality defined this way indeed has the intended property, i.e., equality of two patterns holds iff the two patterns are interpreted to equal sets.
\begin{coqcode}
Lemma equal_iff_interpr_same : forall (M : @Model ?$\Sigma$?),
    M ?$\vDash^T$? theory ->
    forall (?$\phi1$? ?$\phi2$? : Pattern) (?$\rho_e$? : @EVarVal ?$\Sigma$? M) (?$\rho_s$? : @SVarVal ?$\Sigma$? M),
      @eval ?$\Sigma$? M ?$\rho_e$? ?$\rho_s$? (?$\phi$?1 =ml ?$\phi$?2) = ?$\top$? <->
      @eval ?$\Sigma$? M ?$\rho_e$? ?$\rho_s$? ?$\phi1$? = @eval ?$\Sigma$? M ?$\rho_e$? ?$\rho_s$? ?$\phi2$?.
\end{coqcode}

Our formalization also demonstrates that in matching logic, one cannot simply use equivalence ($\leftrightarrow$) instead of equality.
As argued in \cite{Rosu17}, one could expect that the pattern
$\exists \ld f\ x \leftrightarrow \ee{0}$ specifies that $f$ behaves like a function;
however, there exists a model in which that is not the case.
In the model whose domain is \coqinline{exampleDomain} and whose interpretation of
application is defined as \coqinline{example_app_interp} below:
\begin{coqcode}
Inductive exampleDomain : Set := one | two | f.
Definition example_app_interp (d1 d2 : exampleDomain) : Power exampleDomain :=
  match d1, d2 with
  | f, one => ?$\top$?
  | f, two => ?$\emptyset$?
  | _, _ => ?$\emptyset$?
  end.
\end{coqcode}
\noindent the pattern
$\exists.\, f\ x \leftrightarrow \ee{0}$ holds (in every interpretation of $x$),
even though the model does not implement $f$ as a function. For more technical details, we refer to the formalization~\cite{ml-formalization}.

\section{Reasoning in Matching Logic}\label{sec:proofmode}


In~\Cref{sec:formalization}, we encoded matching logic and its proof system in Coq. With the minimal proof system, one can already reason about syntactic consequence by using Coq's \coqinline{apply} tactic.
However, only using the presented rules to reason about derived operations and complex theories is not really productive: it is easy to get lost when facing a complex proof obligation expressed in vanilla matching logic after unfolding the derived notations.

\paragraph{Derived rules.}

To support object-level reasoning, we proved several derived axioms and rules, essentially enriching the proof system with rules about derived constructs (commonly used operations that are not included in the syntax of matching logic) and common theories (such as definedness). These alone can shorten a typical matching logic proof script significantly. For example, think of destructing a disjunctive premise into two premises, which is naturally one step in an informal proof, but takes a couple of steps with the matching logic proof system. However, writing proofs with the derived rules is still cumbersome, because now we get lost in the details of combining and applying these theorems with the correct parameters. We tackle this problem with a dedicated Coq proof mode.

\subsection{Matching Logic Proof Mode}


To further simplify reasoning in the embedded logic, we conceptually separate the matching logic proof state from the Coq proof state, introduce a local proof context, and define a set of special Coq tactics that manipulate the dedicated proof state. We call these concepts together \emph{the matching logic proof mode}\footnote{We borrow the term \emph{proof mode} and the approach from the authors of the Iris proof mode~\cite{IrisProofMode}, and the Coq reference manual~\cite{coq-manual}.}. The ultimate goal with the proof mode is to make matching logic proofs simple to read and write, especially for users familiar with Coq.
The contents of this section are work-in-progress, but nicely demonstrate the potential that lies in carrying out interactive matching logic proofs in Coq.

\paragraph{Matching logic proof state.}

The concept of the proof state allows us to nicely mimic Coq-style reasoning in matching logic by rendering matching logic proof goals as a list of named hypotheses and a goal pattern. Behind the scenes, the goal on provability is turned into a record that stores the list of the premises along with the conclusion.
The proof mode allows for moving left-hand sides of implication conclusions to the list of premises (the local context), which is essential in matching logic as the deduction theorem~\cite{chen-rosu-2019-tr-mml} can be applied to totality patterns only.
The proof mode provides a better overview on the state of the proof in the interactive mode. In particular, it contains the following sections:

\begin{itemize}
\item A meta-level context (such as hypotheses on the well-formedness of patterns)
\item A global matching logic context (a set of patterns known to be valid);
\item A local matching logic context (a named list of patterns assumed to be valid);
\item A matching logic goal (a single pattern, the conclusion).
\end{itemize}

We provide a notation for this proof state, which also resembles the proof state in Coq (see \Cref{fig:proofstate}). In this
example, $\phi_1, \dots, \phi_n$ form the global matching logic context, while $\psi_1, \dots, \psi_n$ form the local one, and $\chi$ is the goal.
A matching logic proof state can automatically be converted to a syntactic provability statement as presented in \Cref{fig:proofstateconv}
which describes this conversion of the proof state in \Cref{fig:proofstate}.


\begin{figure}[h]
  \begin{subfigure}{.49\textwidth}
    \begin{coqcode}
?$\phi_1$?, ?$\dots$?, ?$\phi_n$?, ?$\psi_1$?, ?$\dots$?, ?$\psi_m$?, ?$\chi$? : Pattern
______________________________________(1/1)
{[?$\phi_1, \dots, \phi_n$?]} ?$\vdash$?
"H1" : ?$\psi_1$?
?$...$?
"Hn" : ?$\psi_m$?
--------------------------------------
?$\chi$?
    \end{coqcode}
  \caption{Notation}
  \label{fig:proofstate}
  \end{subfigure}\hfill
  \begin{subfigure}{.49\textwidth}
    \begin{coqcode}
?$\phi_1$?, ?$\dots$?, ?$\phi_n$?, ?$\psi_1$?, ?$\dots$?, ?$\psi_m$?, ?$\chi$? : Pattern
______________________________________(1/1)
{[?$\phi_1, \dots, \phi_n$?]} ?$\vdash$? ?$\psi_1$? ---> ?$\dots$? ---> ?$\psi_m$? ---> ?$\chi$?    
    \end{coqcode}
  \caption{Conversion}
  \label{fig:proofstateconv}
  \end{subfigure}
  \caption{Matching logic proof state}
\end{figure}

The mapping between matching logic proof states and matching logic proofs of syntactic consequence is not injective: there can be multiple proof states that represent the same matching logic proof obligation when the the conclusion is an implication pattern.

\paragraph{Matching logic proof tactics.}

To create proof tactics, we first lift the derived proof rules to work with matching logic proof states. By lifting, 
we actually mean proving the derived rules for the new proof state. The created tactics can be divided into three main groups:

\begin{itemize}
\item Tactics that restructure the local context (e.g., \coqinline{mlIntro}, \coqinline{mlRevertLast}, \coqinline{mlClear}).
\item Tactics that apply lifted derived rules to the proof state (e.g., \coqinline{mlApply}, \coqinline{mlApplyMeta}, \coqinline{mlDestructOr}).
\item Miscellaneous tactics (e.g., \coqinline{mlRewrite} which replaces parts of the matching logic goal, \coqinline{mlTauto} which is a preliminary
tautology solver).
\end{itemize}

We can use these tactics in a similar way as their Coq counterparts (e.g., \coqinline{mlIntro} mimics the effect of \coqinline{intro}), and create matching logic syntactic proofs conveniently. For the sake of brevity, we do not go into details about the implementation of the tactics, but in the background, they expand to applications of the proof system rules, therefore they construct valid matching logic proofs.

\subsection{An Interactive Proof}

In this section, we show an interactive proof outline (\Cref{fig:example}) with the matching logic proof mode. The complete proof is available in the 
formalization~\cite{ml-formalization} (moreover, there is also a short tutorial about the currently formalized tactics). We show an example proof state transformation from each tactic category, but first, we present two lemmas that
are essential to construct the proof.

\begin{figure}[ht]
\begin{coqcode}
?$1.$?  Lemma overlapping_variables_equal {?$\Sigma$? : Signature} {syntax : Syntax} :
?$2.$?    forall x y ?$\Gamma$?, theory ?$\subseteq$? ?$\Gamma$? ->
?$3.$?    ?$\Gamma$? ?$\vdash$? ?$\lceil$? (patt_free_evar y) and (patt_free_evar x) ?$\rceil$? --->
?$4.$?           patt_free_evar y  =ml  patt_free_evar x.
?$5.$?  Proof.
?$6.$?    intros x y ?$\Gamma$? H?$\Gamma$?.
?$7.$?    remember (patt_free_evar x) as pX. assert (well_formed pX) by (rewrite HeqpX;auto).
?$8.$?    remember (patt_free_evar y) as pY. assert (well_formed pY) by (rewrite HeqpY;auto).
?$9.$?    toMLGoal. wf_auto2.
?$10.$?   unfold patt_equal, patt_iff.
?$\textlabel{11.}{l11}$?   mlRewrite (@patt_total_and ?$\Sigma$? syntax ?$\Gamma$?
?$12.$?                                 (pY ---> pX)
?$13.$?                                 (pX ---> pY) H?$\Gamma$?
?$14.$?                                 ltac:(wf_auto2) ltac:(wf_auto2)) at 1.
?$\textlabel{15.}{l15}$?   mlIntro "H0". mlIntro "H1". mlDestructOr "H1" as "H1'" "H1'".
?$\textlabel{16.}{l16}$?.   * mlApply "H1'". mlClear "H1'". mlIntro "H2".
?$17.$?     (* ... *)
?$18.$?   * (* ... *)
?$19.$? Defined. 
\end{coqcode}
\caption{Case Study for the Proof Mode}
\label{fig:example}
\end{figure}

The first lemma is about the connection of conjunction and totality.

\begin{coqcode}
Lemma patt_total_and {?$\Sigma$? : Signature} {syntax : Syntax}:
  forall ?$\Gamma$? ?$\varphi$? ?$\psi$?, theory ?$\subseteq$? ?$\Gamma$? -> well_formed ?$\varphi$? -> well_formed ?$\psi$? ->
  ?$\Gamma$? ?$\vdash$? ?$\lfloor$? ?$\varphi$? and ?$\psi$? ?$\rfloor$? <---> ?$\lfloor$? ?$\varphi$? ?$\rfloor$? and ?$\lfloor$? ?$\psi$? ?$\rfloor$?.
\end{coqcode}

The second lemma is the congruence lemma, which states that one can replace equivalent subpatterns in \emph{any} context results in equivalent patterns.
\begin{coqcode}
Lemma prf_equiv_congruence ?$\Gamma$? p q C:
  PC_wf C -> ?$\Gamma$? ?$\vdash$? (p <---> q) -> ?$\Gamma$? ?$\vdash$? ((C [p]) <---> (C [q])).
\end{coqcode}

We implemented \coqinline{mlRewrite} based on the congruence lemma. At line \ref{l11} (\Cref{fig:example}) we can use this tactic with the first lemma, since it states the equivalence of two patterns. With this, we are able to propagate totality to the subpatterns of the conjunction for our concrete patterns.

\noindent\begin{minipage}{.49\textwidth}
\begin{coqcode}
______________________________________(1/1)
?$\Gamma$? ?$\vdash$?
?$\lceil$? pY and pX ?$\rceil$? ---> 
  ?$\lfloor$? (pY ---> pX) and (pX ---> pY) ?$\rfloor$?
\end{coqcode}
\end{minipage}\hfill
\begin{minipage}{.49\textwidth}
\begin{coqcode}

______________________________________(1/1)
?$\Gamma$? ?$\vdash$?
?$\lceil$? pY and pX ?$\rceil$? --->
  ?$\lfloor$? (pY ---> pX) ?$\rfloor$? and ?$\lfloor$? (pX ---> pY) ?$\rfloor$?
\end{coqcode}
\end{minipage}

Next, in line \ref{l15}, we reshape the structure of the matching logic proof state by using \coqinline{mlIntro} twice. This tactic
moves the left-hand side of the implication in the goal to the local matching logic context (note that conjunction is a syntactic sugar).

\noindent\begin{minipage}{.49\textwidth}
\begin{coqcode}

______________________________________(1/1)
?$\Gamma$? ?$\vdash$?
?$\lceil$? pY and pX ?$\rceil$? ---> 
   ?$\lfloor$? (pY ---> pX) ?$\rfloor$? and ?$\lfloor$? (pX ---> pY) ?$\rfloor$?
\end{coqcode}
\end{minipage}\hfill
\begin{minipage}{.49\textwidth}
\begin{coqcode}
______________________________________(1/1)
?$\Gamma$? ?$\vdash$?
"H0" : ?$\lceil$? pY and pX ?$\rceil$?,
"H1" : ! ?$\lfloor$? pY ---> pX ?$\rfloor$? or ! ?$\lfloor$? pX ---> pY ?$\rfloor$?,
--------------------------------------
?$\bot$?
\end{coqcode}
\end{minipage}

Finally, we also show the usage of \coqinline{mlApply} in line \ref{l16}. The conclusion of \coqinline{"H1'"} matches the goal,
thus we can apply it, and show its premise.

\noindent\begin{minipage}{.49\textwidth}
\begin{coqcode}
______________________________________(1/1)
?$\Gamma$? ?$\vdash$?
"H0" : ?$\lceil$? pY and pX ?$\rceil$?,
"H1'" : ! ?$\lfloor$? pY ---> pX ?$\rfloor$?,
--------------------------------------
?$\bot$?
\end{coqcode}
\end{minipage}\hfill
\begin{minipage}{.49\textwidth}
\begin{coqcode}
______________________________________(1/1)
?$\Gamma$? ?$\vdash$?
"H0" : ?$\lceil$? pY and pX ?$\rceil$?,
"H1'" : ?$\lfloor$? pY ---> pX ?$\rfloor$? ---> ?$\bot$?,
--------------------------------------
?$\lfloor$? pY ---> pX ?$\rfloor$?
\end{coqcode}
\end{minipage}

It can be observed that we used a number of standard Coq tactics, and explicit parameters during the proof (in \Cref{fig:example}). It is ongoing work to continue refining the proof mode and adding new tactics on demand to formalize as many paper-based matching logic proofs in Coq as possible.

\section{Related Work}\label{sec:related-work}

\subsection{Embedding Logical Languages in Coq}\label{sec:embedding}

Ideally, different sorts of problems are specified in different logical languages which fit the problem domain best. For instance, separation logics excel at describing algorithms manipulating shared and mutable states, temporal logics provide abstractions for specifying systems properties qualified in terms of time, whereas matching logic gives a fairly generic basis for reasoning about programming language semantics and program behavior. It is highly desired to carry out proofs in these domain-specific logical systems interactively and mechanically verified, but these logics tend to significantly diverge from the logics of general-purpose proof assistants such as Coq, leading to an abstraction gap.

To use a proof assistant to formalize reasoning in a specific logic, one needs to encode the logic as a theory in the proof assistant and then carry out reasoning at the meta-level with considerable overhead. Related works have been investigating different ways of embedding with the aim of reducing the overhead and facilitate productive object-level reasoning in various logical languages. To name a few, (focused) linear logic~\cite{power1999working,xavier2018mechanizing}, linear temporal logic~\cite{8133459}, different dialects of separation logic~\cite{krebbers2018mosel,mccreight2009practical,appel2007separation} and differential dynamic logic~\cite{10.1145/3018610.3018616} have been addressed in the past. Note that some of these encodings are full-featured proof modes, which create a properly separated proof environment and tactic language for the object logic. A slightly different idea worth mentioning is encoding one theorem prover’s logic in another to make the proofs portable, such as taking HOL proofs to Coq~\cite{wiedijk2007encoding,keller2010importing}.

The existing approaches show significant differences depending upon whether the formalization is aimed at proving the properties of the logic or at advocating reasoning in the logic. One particular consideration is to variable representation and the level of embedding. The majority of the cited formalizations apply a so-called shallow embedding, where they reuse core elements of the meta-logic; for instance, names are realized by using higher-order abstract syntax or the parametrized variant thereof, and exploit the binding and substitution mechanism built-in the proof assistant. With this, name binding, scopes, and substitution come for free, but the formalization is tied to the meta-logic’s semantics in this aspect, which may not be suitable in all cases. In fact, one of the main design decisions in our work was to use deep embedding facilitated by notations and locally nameless variable representation.

\subsection{Matching Logic Implementations}
This paper is not the only attempt that tries to formalize matching logic
using a formal system. 
In \cite{ChenTrustworthyK}, the authors propose a matching logic formalization based on Metamath \cite{metamath}, a formal language used to encode abstract mathematical axioms and theorems. 
The syntax and proof system of matching logic are defined in Metamath
in a few hundreds lines of code \cite{ml-checker}.
Matching logic (meta)theorems can be formally stated in Metamath, and 
their formal proofs can be encoded in Metamath as machine-checkable proof objects.
While the Metamath implementation is simpler with a smaller trust base, the
Coq formalization of matching logic is more versatile; in Coq one can
express a larger variety of metatheorems such as the deduction theorem.
In general, the Metamath formalization focuses only on proofs of explicit
matching logic theories, while our formalization, despite a larger trust-base,
focuses mostly on models, semantics, and metatheorems of matching logic.
This dichotomy allows for potential integration with the Metamath formalization (see~\ref{sec:conclusion}).


Another matching logic implementation is through the \K framework.
The \K framework is a very robust engine for formalizing programming
language syntax and semantics. A version of matching logic, called
Kore, is used by \K to represent processed formal semantics. This is
how full large-scale languages can be simply represented as matching
logic theories. Admittedly, defining matching logic theories on the
scale of programming languages directly in Coq is not currently feasible.
However, our formalization brings more interactivity to matching logic reasoning,
which is currently missing in \K.

\section{Conclusion and Future Work}\label{sec:conclusion}
In this work, we defined a locally nameless representation of matching logic.
We also presented the first formal
definition of \emph{any} version of matching logic using an interactive
theorem prover, namely Coq.
We mechanized the soundness theorem of matching logic, and presented some nontrivial matching logic theories and
interactive proofs with a preliminary matching logic proof mode.
We believe this paves the way for Coq and interactive theorem provers to be
used more frequently with matching logic. We discuss some areas for
future work below.
\begin{itemize}\setlength\itemsep{0em}
\item
\emph{Complete Coq Proof Mode for Matching Logic. }
Proving formulas using the matching logic Hilbert-style proof system is not always convenient,
especially when compared to the way one can prove theorems in Coq.
For this reason we are working on the presented proof mode for matching logic in Coq,
that allows users to prove matching logic theorems using tactics that manipulate the goal
and local context.
We took the inspiration mainly from the Iris project, where the authors built a proof mode
for a variant of separation logic \cite{IrisProofMode}.


\item
\emph{Create Tactics for Type Class Instantiation. } While using the formalization with actual signatures or new derived notations,
the user needs to instantiate certain simple type classes. We plan to create tactics to carry out this work automatically.

\item
\emph{Exporting Metamath Proof Objects. }
An interesting way of combining advantages of both our Coq
formalization and the Metamath formalization
in~\cite{ChenTrustworthyK} would be the ability to convert matching
logic proofs in Coq to matching logic proofs in Metamath.
One challenge here is posed by the fact that Metamath uses the traditional named
representation of matching logic patterns, which is different from the
locally nameless representation used in our Coq development.

\item 
\emph{Importing \K Definitions. }
As mentioned in Section~\ref{sec:related-work}, the \K framework
is a matching logic (specifically Kore) implementation with the
advantage of being able to naturally define real large-scale
programming languages.
As future work, we plan to formalize Kore as a matching logic theory inside Coq
and write a translator from Kore files to Coq files using this theory,
thus giving \K framework a Coq-based backend.
This would allow languages defined in \K and properties of those
languages proved in \K to be \emph{automatically} translated to Coq
definitions and theorems.

\item
\emph{Completeness. }
For the fragment of matching logic without the $\mu$ operator, the proof system is complete.
We would like to formalize the proof of completeness from~\cite{chen-rosu-2019-tr-mml};
however, we expect the proof to be non-constructive, which implies we would not be able
to compute proof terms (and extract Metamath proofs) from proofs of semantic validity.

\end{itemize}

\paragraph{Acknowledgements.} We warmly thank Runtime Verification Inc. for their generous funding support. Supported by the ÚNKP-21-4 New National Excellence Program of the Ministry for Innovation and Technology from the source of the National Research, Development and Innovation Fund.

\pagebreak

\bibliography{bibliography}

\end{document}